# Intelligent Web Agent for Search Engines

Avinash N. Bhute[1], Harsha A. Bhute[2], Dr.B.B.Meshram[3]

*Abstract*

*In this paper we review studies of the growth of the Internet and technologies that are useful for information search and retrieval on the Web. Search engines are retrieve the efficient information. We collected data on the Internet from several different sources, e.g., current as well as projected number of users, hosts, and Web sites. The trends cited by the sources are consistent and point to exponential growth in the past and in the coming decade. Hence it is not surprising that about 85% of Internet users surveyed claim using search engines and search services to find specific information and users are not satisfied with the performance of the current generation of search engines; the slow retrieval speed, communication delays, and poor quality of retrieved results. Web agents, programs acting autonomously on some task, are already present in the form of spiders, crawler, and robots. Agents offer substantial benefits and hazards, and because of this, their development must involve attention to technical details. This paper illustrates the different types of agents ,crawlers, robots ,etc for mining the contents of web in a methodical, automated manner, also discusses the use of crawler to gather specific types of information from Web pages, such as harvesting e-mail addresses.*

## I. INTRODUCTION

The potential Internet-wide impact of autonomous software agents on the World Wide Web [1] has spawned a great deal of discussion and occasional controversy. Based upon experience in the design and operation of the Spider [2], such tools can create substantial value to users of the Web. Unfortunately, agents can also be pests, generating substantial loads on already overloaded servers, and generally increasing Internet backbone traffic. Much of the discussion in this paper is about the agents, its types and architecture.

Avinash N Bhute[1], Author is with Sinhgad College of Engineering, Department of information technology as Assistant Professsor, Pune, Maharastra India. email- anbhute@gmail.com.

Harsha A Bhute[2] is Pursuing ME, Computer Engg. (Computer N/W), S.C.O.E, Pune Maharastra, India. email-harshavi@gmail.com.

Dr.B.B.Meshram[3] is professor and head of computer engineering department. VJTI, Mumbai, Maharastra, India email-bbmeshram@vjti.ac.in

## II. WEB CRAWLER

A Web crawler is a computer program that browses the World Wide Web in a methodical, automated manner. Other terms for Web crawlers are ants, automatic indexers, bots, and worm [3] or Web spider, Web robot. Web crawling or spidering is term alternatively used for same. Many sites, in particular search engines, use spidering as a means of providing up-to-date data. Web crawlers are mainly used to create a copy of all the visited pages for later processing by a search engine that will index the downloaded pages to provide fast searches. Crawlers can also be used for automating maintenance tasks on a Web site, such as checking links or validating HTML code. Also, crawlers can be used to gather specific types of information from Web pages, such as harvesting e-mail addresses (usually for spam). A Web crawler is one type of bot, or software agent. In general, it starts with a list of URLs to visit, called the seeds. As the crawler visits these URLs, it identifies all the hyperlinks in the page and adds them to the list of URLs to visit, called the crawl frontier. URLs from the frontier are recursively visited according to a set of policies.

### A. Crawling policies

There are important characteristics of the Web that make crawling it very difficult:
- Large volume,
- Fast rate of change, and
- Dynamic page generation

These characteristics combine to produce a wide variety of possible crawlable pages.

The large volume implies that the crawler can only download a fraction of the Web pages within a given time. The behavior of a Web crawler is the outcome of a combination of following policies:
- Selection Policy
- Re-Visit Policy
- Politeness Policy
- Parallelization Policy

*i. Selection policy*

It states which pages to download, As a crawler always downloads just a fraction of the Web pages, The current size of the Web, evens large search engines cover only a portion of the publicly-available Internet; a study by Lawrence and Giles showed that no search engine indexes more than 16% of the Web in 1999[4].It is highly desirable that the downloaded fraction contains the most relevant pages and not just a random sample of the Web. Cho and other author. made the first study on policies for crawling scheduling. Their data set was a 180,000-pages crawl from the stanford.edu domain, in which a crawling simulation was done with different strategies.[5] The ordering metrics tested were breadth-first, backlink-count and partial Pagerank calculations. One of the conclusions was that if the crawler wants to download pages with high Pagerank early during the crawling process, then the partial Pagerank strategy is the better, followed by breadth-first and backlink-count. However, these results are for just a single domain.

Najork and Wiener [6] performed an actual crawl on 328 million pages, using breadth-first ordering. They found that a breadth-first crawl captures pages with high Pagerank early in the crawl (but they did not compare this strategy against other strategies). The explanation given by the authors for this result is that "the most important pages have many links to them from numerous hosts, and those links will be found early, regardless of on which host or page the crawl originates.

*ii. Re-visit Policy*

It states that to check for changes to the pages. The Web has a very dynamic nature, and crawling a fraction of the Web can take a really long time, usually measured in weeks or months. By the time a Web crawler has finished its crawl, many events could have happened. These events can include creations, updates and deletions.

From the search engine's point of view, there is a cost associated with not detecting an event, and thus having an outdated copy of a resource. The most-used cost functions are freshness and age[7]. Two simple re-visiting policies were studied by Cho and Garcia-Molina[8]:

• Uniform policy: This involves re-visiting all pages in the collection with the same frequency, regardless of their rates of change.

• Proportional policy: This involves re-visiting more often the pages that change more frequently. The visiting frequency is directly proportional to the (estimated) change frequency.

(In both cases, the repeated crawling order of pages can be done either in a random or a fixed order.)

Cho and Garcia-Molina proved the surprising result that, in terms of average freshness, the uniform policy outperforms the proportional policy in both a simulated Web and a real Web crawl. The explanation for this result comes from the fact that, when a page changes too often, the crawler will waste time by trying to re-crawl it too fast and still will not be able to keep its copy of the page fresh

*iii. Politeness Policy*

It states how to avoid overloading Web sites. Crawlers can retrieve data much quicker and in greater depth than human searchers, so they can have a crippling impact on the performance of a site. Needless to say, if a single crawler is performing multiple requests per second and/or downloading large files, a server would have a hard time keeping up with requests from multiple crawlers.

As noted by Koster[9], the use of Web crawlers is useful for a number of tasks, but comes with a price for the general community. The costs of using Web crawlers include:

• Network resources, as crawlers require considerable bandwidth and operate with a high degree of parallelism during a long period of time;

• Server overload, especially if the frequency of accesses to a given server is too high;

• Poorly-written crawlers, which can crash servers or routers, or which download pages they cannot handle; and

• Personal crawlers that, if deployed by too many users, can disrupt networks and Web servers.

A partial solution to these problems is the robots_exclusion protocol, also known as the robots.txt protocol[10] that is a standard for administrators to indicate which parts of their Web servers should not be accessed by crawlers. This standard does not include a suggestion for the interval of visits to the same server, even though this interval is the most effective way of avoiding server overload. Recently commercial search engines like Ask_Jeeves, MSN and Yahoo are able to use an extra "Crawl-delay:" parameter in the robots.txt file to indicate the number of seconds to delay between requests.

*iv. Parallelization Policy*

It states how to coordinate distributed Web crawlers. A parallel crawler is a crawler that runs multiple processes in parallel. The goal is to maximize the download rate while minimizing the overhead from parallelization and to avoid repeated downloads of the same page. To avoid downloading the same page more than once, the crawling system requires a policy for assigning the new URLs discovered during the crawling process, as the same URL can be found by two different crawling processes.

III. WEB CRAWLER ARCHITECTURE

Web crawlers are a central part of search engines, and details on their algorithms and architecture are kept as business secrets. When crawler designs are published, there is often an important lack of detail that prevents others from reproducing the work. There are also emerging concerns about "search_engine_spamming", which prevent major search engines from publishing their ranking algorithms.

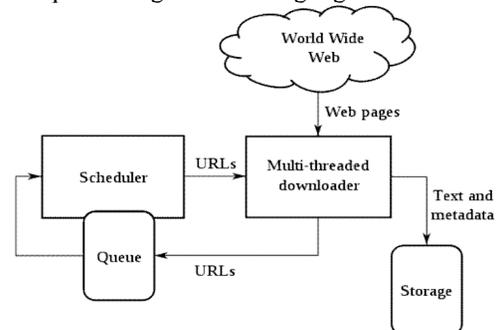

Fig1. Web Crawler Architecture [courtesy ww.linux.ime/WebCrawler.html ]

## IV. EXAMPLES OF WEB CRAWLERS

The following is a list of published crawler architectures for general-purpose crawlers (excluding focused web crawlers), with a brief description that includes the names given to the different components and outstanding features:

- Yahoo Crawler (Slurp) is the name of the Yahoo Search crawler.
- FAST Crawler [11]is a distributed crawler, used by Fast Search & Transfer,
- Google Crawler The crawler was integrated with the indexing process, because text parsing was done for full-text indexing and also for URL extraction. There is a URL server that sends lists of URLs to be fetched by several crawling processes. During parsing, the URLs found were passed to a URL server that checked if the URL have been previously seen. If not, the URL was added to the queue of the URL server.
- Methabot is a scriptable web crawler written in C, released under the ISC license.
- PolyBot is a distributed crawler, which is composed of a "crawl manager", one or more "downloaders" and one or more "DNS resolvers". Collected URLs are added to a queue on disk, and processed later to search for seen URLs in batch mode.
- RBSE was the first published web crawler. It was based on two programs:"spider" maintains a queue in a relational database, and "mite", is a modified www ASCII browser that downloads the pages from the Web.
- WebCrawler was used to build the first publicly-available full-text index of a subset of the Web. It was based on lib-WWW to download pages, and another program to parse and order URLs for breadth-first exploration of the Web graph. It also included a real-time crawler that followed links based on the similarity of the anchor text with the provided query.
- World Wide Web Worm was a crawler used to build a simple index of document titles and URLs. The index could be searched by using the grep Unix command.
- WebFountain is a distributed, modular crawler similar to Mercator. It features a "controller" machine that coordinates a series of "ant" machines. After repeatedly downloading pages, a change rate is inferred for each page and a non-linear programming method must be used to solve the equation system for maximizing freshness.
- WebRACE is a crawling and caching module, used as a part of a more generic system called eRACE. The system receives requests from users for downloading web pages, so the crawler acts in part as a smart proxy server. The system also handles requests for "subscriptions" to Web pages that must be monitored: when the pages change, they must be downloaded by the crawler and the subscriber must be notified. The most outstanding feature of WebRACE is that, while most crawlers start with a set of "seed" URLs, WebRACE is continuously receiving new starting URLs to crawl from.

In addition to the specific crawler architectures listed above, there are general crawler architectures published by Cho[12] and Chakrabarti.[13]

**Open-source crawlers**
- Aspseek is a crawler, indexer and a search engine.
- DataparkSearch is a crawler and search engine released under the GNU General Public License.
- GNU Wget is a command-line-operated crawler. It is typically used to mirror Web and FTP sites.
- GRUB is an open source distributed search crawler that Wikia Search ( http://wikiasearch.com ) uses to crawl the web.
- Heritrix is the Internet Archive's archival-quality crawler, designed for archiving periodic snapshots of a large portion of the Web. It was written in Java.
- HTTrack uses a Web crawler to create a mirror of a web site for off-line viewing. It is written in C and released under the GPL.
- ICDL Crawler is a cross-platform web crawler, intended to crawl Web sites based on Web-site Parse Templates using computer's free CPU resources only.
- mnoGoSearch is a crawler, indexer and a search engine.
- Nutch is a crawler can be used in conjunction with the Lucene text-indexing package.
- Pavuk is a command-line Web mirror tool with optional X11 GUI crawler,It has bunch of advanced features compared to wget and httrack, e.g., regular expression based filtering and file creation rules.
- YaCy, a free distributed search engine, built on principles of peer-to-peer networks.

## V. WEB ROBOT

Internet bots, also known as web robots, simply bots, are software applications that run automated tasks over the Internet. Typically, bots perform tasks that are both simple and structurally repetitive, at a much higher rate than would be possible for a human alone. The largest use of bots is in web spidering, in which an automated script fetches, analyses and files information from web servers at many times the speed of a human. Each server can have a file called robots.txt, containing rules for the spidering of that server that the bot is supposed to obey. In addition to their uses outlined above, bots may also be implemented where a response speed faster than that of humans is required (*e.g.*, gaming bots and auction-site robots) or less commonly in situations where the emulation of human activity is required, i.e. chat bots. These chatterbots may allow people to ask in plain English and then formulate a proper response. These bots can often handle many tasks, including reporting weather, zip-code information, sports scores, converting currency or other units, etc. Others are used for entertainment, such as SmarterChild on AOL Instant Messenger and MSN Messenger and Jabberwacky on Yahoo! Messenger

A. A Chatterbot

Chatterbot (or chatbot)[15],[16] is a type of conversational agent, a computer program designed to simulate an intelligent conversation with one or more human users via auditory or textual methods. Such computer programmes can also referred to as Artificial Conversational Entities within some specific contexts. Although many chatterbots appear to intelligently interpret human input prior to providing a response, many simply scan for keywords within the input and pull a reply with the most matching keywords or the most similar wording pattern from a local database.

B. Chatterbot Development

The classic historic early chatterbots are ELIZA (1966) and PARRY (1972).[17],[18] More recent notable programs include A.L.I.C.E. and Jabberwacky.

The growth of chatterbots as a field of research has created an expansion in their range of purposes and potential applications. While ELIZA and PARRY were used exclusively to simulate typed conversation, many now include functional features such as games and web searching abilities. A program called Racter has also been used to "write" an original story called *The Policeman's Beard is Half Constructed*.

One pertinent field of AI research is natural language processing. Usually, weak AI fields employ specialized software or programming languages created specifically for the narrow function required. For example, A.L.I.C.E., utilises a programming language called AIML which is specific to its function as a conversational agent, and has since been adopted by various other developers of, so called, Alicebots. Nevertheless, A.L.I.C.E. is still purely based on pattern matching techniques without any reasoning capabilities, and this is distinguished from strong AI, which would require sapience and logical reasoning abilities. This is the same technique ELIZA, the first chatterbot, was using back in 1966.

Another notable program, known as Jabberwacky, learns new responses and context based on real-time user interactions, rather than being driven from a static database like many other existing chatterbots. Some more recent chatterbots also combine real-time learning with evolutionary algorithms which optimise their ability to communicate based on each conversation held, with one notable example being Kyle, winner of the 2009 Leodis AI Award.[1] Although such programs show initial promise, many of the existing results in trying to tackle the problem of natural language still appear fairly poor, and it seems reasonable to state that there is currently no general purpose conversational artificial intelligence. This has led some software developers to focus more on the practical aspect of chatterbot technology - information retrieval.

A common rebuttal often used within the AI community against criticism of such approaches asks, "How do we know that humans don't also just follow some cleverly devised rules?" (in the way that Chatterbots do). Two famous examples of this line of argument against the rationale for the basis of the Turing test are John Searle's Chinese room argument and Ned Block's Blockhead argument.

C. Example of Chatterboats
- Classic chatterbots
- Dr. Sbaitso
- ELIZA
- PARRY
- Racter

General chatterbots
- A.L.I.C.E., 2001, 2002, and 2004 Loebner Prize winner developed by Richard Wallace.
- Albert One, 1998 and 1999 Loebner winner, by Robby Garner
- Claude, or Racter, released 1983
- Elbot, 2008 Loebner Prize winner, by Fred Roberts
- Fred, an early chatterbot by Robby Garner
- Jabberwacky
- Jeeney AI
- MegaHal
- Spookitalk - A chatterbot used for NPCs in Douglas Adams' *Starship Titanic* video game.
- Ultra Hal Assistant
- Verbot

VI. CONCLUSION

Intelligent agents are an essential component of all search engines, and are increasingly becoming important in data mining and other indexing applications. While some have questioned whether such exponential growth is currently being maintained. The trend towards automated production of web pages from databases makes it likely that such growth will continue, or even accelerate, in the immediate future. Intelligent search agents like web crawler efficiently crawl the web and providing best result which is scalable and efficient.

REFERENCES

[1] Berners-Lee, T., R. Cailliau, A. Loutonen, H. F. Nielsen and A. Secret, "The World- Wide Web," Communications of the ACM, v. 37, n. 8, August 1994, p. 76-82.

[2] Eichmann, D. "The RBSE Spider -- Balancing Effective Search against Web Load," First International Conference on the World Wide Web, Geneva, Switzerland, May 25-27, 1994, p. 113-120.

[3] Kobayashi, M. and Takeda, K. (2000). "Information retrieval on the web". ACM Computing Surveys (ACM Press) **32** (2): 144–173. doi:10.1145/358923.358934. http://doi.acm.org/10.1145/358923.358934

[4] Lawrence, Steve; C. Lee Giles (1999-07-08). "Accessibility of information on the web". Nature **400** (6740): 107. doi:10.1038/21987.

[5] Cho, J.; Garcia-Molina, H.; Page, L. (1998-04), "Efficient Crawling Through URL Ordering", Seventh International World-Wide Web Conference, Brisbane, Australia

[6] Marc Najork and Janet L. Wiener. Breadth-first crawling yields high-quality pages. In Proceedings of the Tenth Conference on World Wide Web, pages 114–118, Hong Kong, May 2001. Elsevier Science.

[7] Cho, Junghoo; Hector Garcia-Molina (2000). "Synchronizing a database to improve freshness". Proceedings of the 2000 ACM SIGMOD international conference on Management of data. Dallas, Texas, United States: ACM. pp. 117-128. doi:10.1145/342009.335391. ISBN 1-58113-217-4.

[8] Cho, J. and Garcia-Molina, H. (2003). Effective page refresh policies for web crawlers. ACM Transactions on Database Systems, 28(4).

[9] Koster, M. (1995). Robots in the web: threat or treat ? ConneXions, 9(4).

[10] Koster, M. (1996). A standard for robot exclusion.

[11] K. M. and Michelsen, R. (2002). Search Engines and Web Dynamics. Computer Networks, vol. 39, pp. 289–302, June 2002.

[12] Cho, Junghoo; Hector Garcia-Molina (2002). "Parallel crawlers". Proceedings of the 11th international conference on World Wide Web. Honolulu, Hawaii, USA: ACM. pp. 124-135. doi:10.1145/511446.511464. ISBN 1-58113-449-5.

[13] Chakrabarti, S. (2003). Mining the Web. Morgan Kaufmann Publishers. ISBN 1-55860-754-4

[14] Güzeldere, Güven; Franchi, Stefano (1995-07-24), dialogues with colorful personalities of early ai, "Constructions of the Mind", Stanford Humanities Review, SEHR (Stanford University), http://www.stanford.edu/group/SHR/4-2/text/dialogues.html, retrieved 2008-03-05

[15] Mauldin, Michael (1994), "ChatterBots, TinyMuds, and the Turing Test: Entering the Loebner Prize Competition", Proceedings of the Eleventh National Conference on Artificial IntelligenceAAAIPress, http://www.aaai.org/ Library/AAAI/aaai94contents.php, retrieved 2008-03-05

[16] Network Working Group (1973), "RFC 439, PARRY Encounters the DOCTOR", The Internet Engineering Task Force (Internet Society), http://tools.ietf.org/html/rfc439, retrieved 2008-03-05

[17] Sondheim, Alan J (1997), <nettime> Important Documents from the Early Internet (1972),nettime.org, http://www.nettime.org/ Lists-Archives/nettime-l-9707/msg00059.html, retrieved 2008-03-05